# Extinction and re-initiation of methane detonation in dilute coal particle suspensions

Jingtai Shi[a,b], Yong Xu[b], Pikai Zhang[b,c], Wanxing Ren[a], Huangwei Zhang[b,*]

[a] *School of Safety Engineering, China University of Mining and Technology, Xuzhou, 221116, China*
[b] *Department of Mechanical Engineering, National University of Singapore, 9 Engineering Drive 1, Singapore 117576, Republic of Singapore*
[c] *National University of Singapore (Chongqing) Research Institute, Liangjiang New Area, Chongqing 401123, China*

___________________________________________________________________________

**Abstract**

Methane/coal dust hybrid explosion is one of the common hazards in process and mining industries. In this study, methane detonation propagation in dilute coal particle suspensions is studied based on Eulerian-Lagrangian method. Two-dimensional configuration is considered, and a skeletal chemical mechanism (24 species and 104 reactions) is applied for methane combustion. The gas and particulate phase equations are solved using an OpenFOAM code for two-phase compressible reacting flow, RYrhoCentralFOAM. The effects of char combustion on methane detonation dynamics are investigated and devolatized coal particles are modelled. The results show that propagation of the methane detonation wave in coal particle suspensions are considerably affected by coal particle concentration and size. Detonation extinction occurs when the coal particle size is small and concentration is high. The averaged lead shock speed generally decreases with increased particle concentration and decreased particle size. Mean structure of methane and coal particle hybrid detonation is analysed, based on the gas and particle quantities. It is found that char combustion proceeds in the subsonic region behind the detonation wave and heat release is relatively distributed compared to that from gas phase reaction. Moreover, for 1 μm particle, if the particle concentration is beyond a threshold value, detonation re-initiation occurs after it is quenched at the beginning of the coal dust suspensions. This is caused by hot spots from the shock focusing along the reaction front in a decoupled detonation and these shocks are generated from char combustion behind the lead shock. A regime map of detonation propagation and extinction is predicted. It is found that the re-initiation location decreases with the particle concentration and approaches a constant value when the concentration exceeds 1000 g/m$^3$. The results from this study are useful for prevention and suppression of methane/coal dust hybrid explosion.

*Keywords:* Detonation extinction; re-initiation; methane; coal particle; char combustion

___________________________________________________________________________

*Corresponding author. E-mail address:
huangwei.zhang@nus.edu.sg (H. Zhang)

## 1. Introduction

Methane/coal dust hybrid explosion is one of the common hazards in process and mining industries [1]. After being heated by hot surrounding gas where the dust particles are dispersed, their devolatilization and/or surface reaction can be initiated, through which volatile gas and reaction heat are released. This would considerably modulate the thermodynamic state of local flammable gas (e.g., methane/air mixture). Typically, existence of coal dust would complicate a gas explosion process and therefore make it more difficult to be predicted [1]. Due to harsh experimental conditions and demanding requirement for modelling strategies to reproduce the multi-faceted physics, our understanding about hybrid explosion in methane and coal dust mixtures is still rather limited.

Investigations have been available about flammability limit, ignitability, and flame propagation in methane/coal dust two-phase mixtures. For instance, Cloney et al. [2] investigated the burning velocity and flame structures of hybrid mixtures of coal dust with methane below the lower flammability limit of the gaseous mixture. They correlated the unsteady flame behaviors (e.g., burning velocity oscillation) with combustion of volatile gas released from the dispersed particles. Xu et al. [3] found that both maximum explosion overpressure and overpressure rise rate increase with increased coal dust concentrations and decreased diameter. Xu et al. also studied the performance of mitigation of methane/coal dust explosion with fine water sprays [3,4]. Xie et al. [5] observed that flame burning velocity decreases when coal particle of sizes 53-63 μm and 75-90 μm are added, irrespective of the gas equivalence ratios. They also identified two competing effects associated with the volatile gas release (heat absorption, as thermal effect) and addition (kinetic effect). Rockwell and Rangwala [6] found that turbulent burning velocity of methane flames increases as the coal particle size decreases and the concentration increases (>50%). This is in line with the findings by Chen [7], where he observed that presence of methane in coal dust explosions enhances the flame velocity of the mixture. Furthermore, Amyotte et al. [8] studied the ignitability of methane/coal dust mixture and found that the apparent lean flammability limit decreases with high methane concentration, small particle diameter, and increased volatile matter content.

Houim et al. [9] studied the layered coal dust combustion induced by a blast wave degraded from a methane detonation. It is shown that the high-speed post-shock flow lifts the coal dust at the bottom of the domain, which ignites by a reaction wave of burning carbon char and generates a shock-flame complex. The coal-dust combustion generates pressure waves that overtake the lead shock and intensify the latter. In a subsequent study [10], they also found that inert layers of dust substantially reduce the overpressure, impulse, and speed produced by the propagating blast wave. The shock and flame are more strongly coupled for loose dust layers (initial volume fraction 1%), thereby propagating at a higher velocity and producing large overpressures. More recently, Guhathakurta and Houim observed that the role of heat radiation in layered dust explosions is affected by coal dust volume fraction [11]. With the similar configuration, Shimura et al. [12] investigated the flame structure during methane shock-wave-induced layered coal dust combustion. They found that the dust particles mainly devolatilize behind the reaction front. In the above work [9–12], since only incident blast wave is considered, how methane detonation interacts with the coal dust is not still clear. Moreover, for micro-sized coal dust, they may be easily aerosolized in the air by any aerodynamic perturbation. Therefore, it is necessary to understand how the coal dust suspensions affect an incident propagating detonation wave.

In this study, detonation in methane and coal dust two-phase mixtures will be simulated based on Eulerian-Lagrangian method. The effects of coal particle concentration and size on methane detonation dynamics will be analyzed. The objectives of this work are to study: (1) the effects of coal particle suspensions on methane detonation dynamics; (2) detailed methane/coal particle hybrid detonation structures; (3) mechanism of detonation extinction and re-initiation in coal particle suspensions.

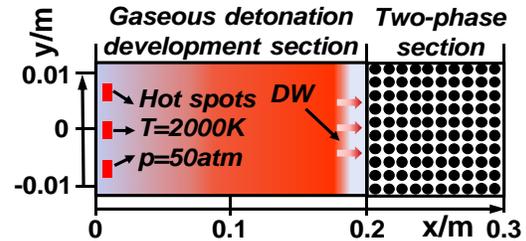

Fig. 1. Schematic of the physical problem. $x$ and $y$ axes not to scale.

## 2. Physical problem

Propagation of an incident methane detonation wave in dilute coal particle suspensions will be studied based on a two-dimensional configuration. The schematic of the physical model is shown in Fig. 1. The length ($x$-direction) and width ($y$) of the domain are 0.3 m and 0.025 m, respectively. It includes gaseous detonation development section (0−0.2 m) and gas-particulate two-phase section (0.2−0.3 m). The whole domain is initially filled with stoichiometric $CH_4/O_2/N_2$ (1:2:1.88 by vol.) mixture. The initial gas temperature and pressure are $T_0$ = 300 K and $p_0$ = 50 kPa, respectively. In the two-phase section, coal particles are uniformly distributed, to mimic coal dust suspensions in methane explosion hazards. In this study, the coal particle diameter varies from $d_p^0 = 1$ to 10 μm. The coal particle concentration ranges from $c$ = 10 to 1000 g/m$^3$. The resultant initial volume fractions are 0.0007%-0.067%, which are well below the upper limiting volume fraction (0.1%) for dilute particle-laden flows [13]. In our simulations, devolatilized coal particles are considered. This enables us to concentrate on the effects of char combustion (interchangeably heterogeneous surface reaction) on methane detonation dynamics. The particle is composed of inert ash and fixed carbon, with mass fractions of 11.28% and 88.72%, respectively. The heat capacity and material density of the particle are 710 J kg$^{-1}$ K$^{-1}$ and 1,500 kg m$^{-3}$. These properties approximately follow those of typical bituminous coal.

The detonation wave (DW) is initiated by three vertically placed hot spots (2 mm × 4 mm, 2,000 K and 50 atm) at the left end of the domain (see Fig. 1). The detonation development section is sufficiently long to achieve a freely propagating methane detonation wave. For all gas-particulate detonations simulations, a consistent initial field with propagating detonation wave at about $x$ = 0.196 m (i.e., slightly before the two-phase section) is used. Upper and lower boundaries of the domain in Fig. 1 are periodic. At $x$ = 0, non-reflective condition is enforced for the pressure, whereas zero-gradient condition for other quantities. At $x$ = 0.3 m, zero-gradient conditions are employed for all variables. Cartesian cells are used to discretize the domain in Fig. 1 and the mesh size is 50 μm at $x$ = 0–0.14 m and 25 μm at 0.14–0.3 m. The resultant cell numbers are 7,800,000. We also perform the mesh sensitivity analysis through halving the resolution in the two-phase section (see supplementary document), which shows that the detonation cell regularity and size are generally close. The same mesh is used in our recent work on methane detonation inhibition by fine water mists [14].

## 3. Mathematical model

The Eulerian-Lagrangian method is used for simulating methane/coal particle hybrid detonations. For the gas phase, the Navier-Stokes equations are solved for the multi-species, compressible, reacting flows. A reduced methane mechanism (DRM 22) [15] is used, including 24 species and 104 reactions. Its accuracy in predicting ignition delay and detonation properties is confirmed in supplementary document through comparisons with detailed chemistry.

In the particulate phase, the Lagrangian method is used to track coal particles. Particle collisions are neglected because of their dilute concentrations. It is assumed that the temperature is uniform inside the particles due to their low Biot numbers (<0.0046). Gravitational force is not included due to smallness of the particles. Coal particles are assumed to spherical, and the swelling effect is not considered. Therefore, the particle size is constant throughout our simulations. With above assumptions, the evolutions of mass, momentum, and energy of a particle are governed by

$$\frac{dm_p}{dt} = -\dot{m}_p, \quad (1)$$

$$m_p \frac{d\mathbf{u}_p}{dt} = \mathbf{F}_d + \mathbf{F}_p, \quad (2)$$

$$m_p c_{p,p} \frac{dT_p}{dt} = \dot{Q}_s + \dot{Q}_c - Q_{p,rad} + Q_{g,rad-p}, \quad (3)$$

where $t$ is time, $m_p = \pi \rho_p d_p^3/6$ is the mass of a single particle, $\rho_p$ and $d_p$ are the particle material density and diameter, respectively. $\dot{m}_p$ is the surface reaction rate and $\mathbf{u}_p$ is the particle velocity vector. The Stokes drag force is modelled as $\mathbf{F}_d = (18\mu/\rho_p d_p^2)(C_d Re_p/24) m_p (\mathbf{u} - \mathbf{u}_p)$, while the pressure gradient force is $\mathbf{F}_p = -V_p \nabla p$. $p$ is pressure, $Re_p$ the particle Reynolds number, $C_d$ the drag coefficient, $\mathbf{u}$ the gas velocity vector, and $V_p$ the single particle volume. In Eq. (3), $c_{p,p}$ is the particle heat capacity and $T_p$ is the particle temperature. $\dot{Q}_s$ is the rate of char combustion heat release absorbed by the particle. The convective heat transfer rate is $\dot{Q}_c = h_c A_p (T - T_p)$, where $h_c$ is the convective heat transfer coefficient, $A_p$ is the particle surface area, $T$ is the gas temperature. Moreover, the radiative emission rate from a particle reads $Q_{p,rad} = A_p \varepsilon_p \sigma T_p^4$, where $\varepsilon_p$ is the emissivity of particle surface and is assumed to unity because the major composition is carbon [16]. $\sigma$ is the Stephen–Boltzmann constant. The particle radiation absorption rate is calculated from $Q_{g,rad-p} = A_p \varepsilon_p \int_{4\pi} I \, d\Omega/4$. $\Omega$ is the solid angle. $I$ is the radiation intensity, solved from the radiative transfer equation using discrete ordinate method.

The char combustion follows a global reaction, C$_{(s)}$ + O$_2$ → CO$_2$, where C$_{(s)}$ is fixed carbon. The kinetic/diffusion-limited rate model [17] is used to estimate the reaction rate, i.e., $\dot{m}_p = A_p p_{ox} D_0 R_k / (D_0 + R_k)$, which accounts for the particle mass change in Eq. (1). $p_{ox}$ is the partial pressure of oxidant species in the surrounding gas. The diffusion rate coefficient $D_0$ and kinetic rate coefficient $R_k$ are respectively estimated from $D_0 = C_1 [(T + T_p)/2]^{0.75}/d_p$ and $R_k = C_2 e^{-(E/RT_p)}$. The constants $C_1$ and $C_2$ are 5×10$^{-12}$ kg/(m·s·Pa·K$^{0.75}$) and 0.002 kg/(m$^2$·s·Pa), respectively, whilst the activation energy $E$ is 7.9×10$^7$ J/kmol [18,19].

The gas and particulate phase equations are solved using an OpenFOAM code for two-phase reacting flow, *RYrhoCentralFOAM* [20–24]. For the gas phase equations, second-order backward scheme is employed for temporal discretization and the time step is about 9×10$^{-10}$ s. A MUSCL-type scheme [25] with van Leer limiter is used for convective flux calculations in momentum equations. Total variation diminishing scheme is used for the convection terms in energy and species equations. Second-order central differencing is applied for the diffusion terms. For the solid phase, Eqs. (1)-(3) are integrated with Euler implicit method and the right-side terms are treated in a semi-implicit fashion. Computational parcel concept is used. The parcel number in our simulations is about 5 million, and the coal particle number in a parcel is specified following the particle size and concentration. Details about the numerical methods in *RYrhoCentralFOAM* can be found in Refs. [21,26].

Validations and verifications of *RYrhoCentralFOAM* have been detailed in [21,22], e.g., shock-capturing, molecular diffusion, flame-chemistry interactions, and two-phase coupling. Here we further validate the solver for shock-particle interactions against the experiments by Sommerfeld [27]. A shock wave of Mach 1.49 propagates into a particle-laden area, and the particles are spherical glass beads. The particle material density, heat capacity and mean diameter are 2.5 g/cm$^3$, 766 J/kg/K, and 27 µm, respectively. Figure 2 shows that our solver can accurately reproduce the evolutions of the shock Mach number for different initial particle volume fractions $\alpha_p$. This further corroborates the solver accuracy for predicting particulate flows.

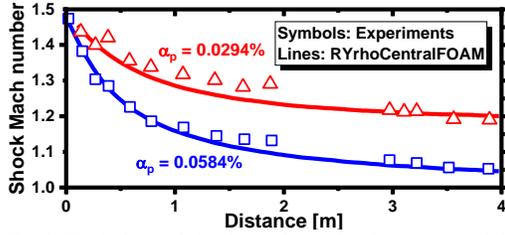

Fig. 2. Evolutions of shock Mach number in shock-particle interactions. Experimental data: Ref. [27].

## 4. Results and discussion

*4.1 DW propagation in coal particle suspensions*

Figure 3 shows the peak pressure trajectories for methane detonations with various coal particle concentrations (10-1000 g/m$^3$). The particle diameter is $d_p = 1$ µm. Evident from Fig. 3 is that coal particle suspensions considerably change the cellular structures of methane detonations. Specifically, when $c = 10$ g/m$^3$, the cells become more regular and the size is smaller, which are particularly pronounced for $x > 0.25$ m, through comparisons with the particle-free case ($c = 0$) in Fig. 3(a). This indicates more stable propagation occurs when relatively light particles are loaded. This is because the combustible coal particles provide an additional heat release, generating shock impulse towards the lead shock and hence enhancing the frontal stability[28]. When $c$ is further increased, e.g., 50 and 250 g/m$^3$, the cell size generally increases, with the mean cell widths $\bar{\lambda}$ of 5.6 and 12.5 mm, respectively.

However, when $c = 500$ and 1000 g/m$^3$, the DW extinction occurs when it immediately encroaches the coal particle area, featured by gradually decreasing overpressures. This is because the reactivity at the triple points (where the trajectories are mostly from) along the DW is highly reduced due to the decoupling of reactive front and lead shock. Nonetheless, interestingly, the detonations are re-initiated downstream in the particle suspensions. This is accompanied by sudden intensification of local peak pressures, as marked as discrete high-pressure spots (HPS) in both Figs. 3(e) and 3(f). The transient and mechanism of DW re-initiation in coal particle suspensions will be further analyzed in Section 4.3, and the unsteady evolutions can be watched in the videos submitted with this manuscript. Further downstream, clear detonation cells appear again, but the strength of the trajectories differs compared to those, e.g., in Fig. 3(d). This is more notable in Fig. 3(f). For instance, in cell A, the weak trajectories are highlighted with dashed lines, which is different from pure gas phase and two-phase steadily propagating detonations in Figs. 3(a)-3(d). This is because the reaction fronts behind the Mach stem are decoupled. After the two triple points (TP1 and TP2) collide, a new Mach stem is formed, and the pressure peak trajectory is strengthened (solid edges of cell A).

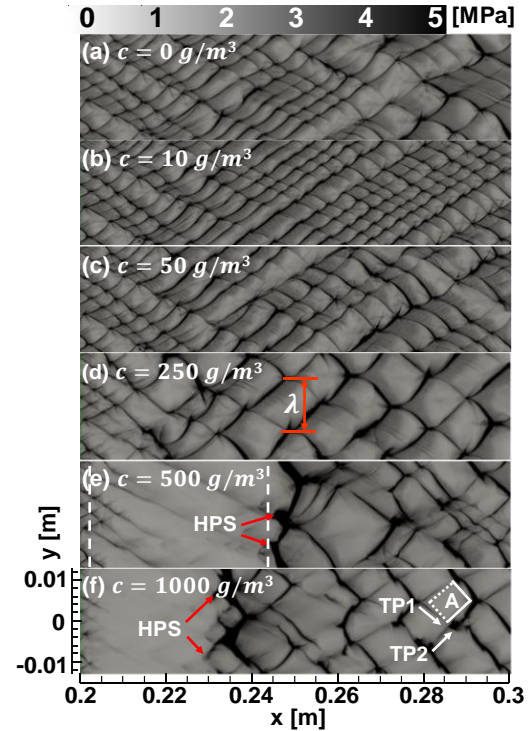

Fig. 3. Peak pressure trajectories with different coal particle concentrations. $d_p = 1$ µm. TP: triple point; HPS: high-pressure spot.

Figure 4 shows the evolutions of lead shock propagation speed $D$ in the two-phase section (0.2-0.3 m) for six cases in Fig. 3. Note that they are calculated from the time history of lead shock positions with a time interval of one microsecond. As demonstrated from lines b and c, $D$ fluctuates around $D_{CJ}$, the Chapman–Jouguet (CJ) speed of the particle-free $CH_4/O_2/N_2$ mixture. This is like the results of gas-only case in Fig. 4(a). However, with $c = 250$ g/m$^3$, the DW has generally lower and more fluctuating speed. This is caused by stronger energy absorption and momentum exchange by more coal particles. For cases e and f, the lead shock speed is considerably reduced to around 70% and 55% of $D_{CJ}$, respectively, before $x > 0.24$ m. This can be justified by the decoupling of reactive front from the lead shock wave, as evidenced in Figs. 3(e) and 3(f). Nonetheless, for $x \geq 0.24$ m, since re-initiation occurs, the lead shock speeds are quickly restored, but still well below the CJ speed.

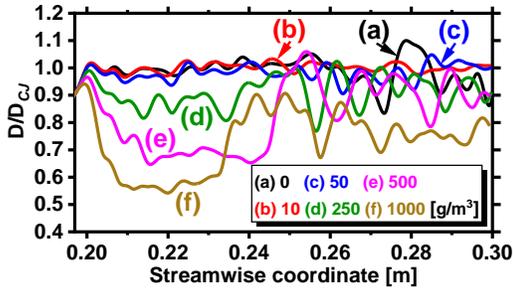

Fig. 4. Evolutions of lead shock speed with various coal particle concentrations. $D_{CJ}$ is the Chapman–Jouguet speed (2,109 m/s) for particle-free $CH_4/O_2/N_2$ mixture.

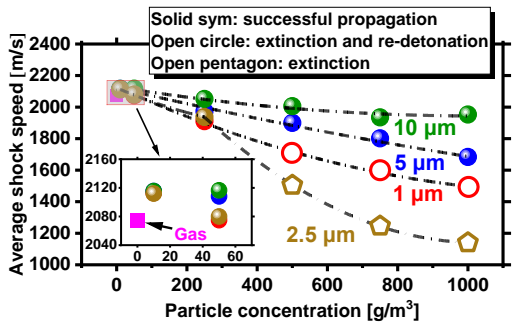

Fig. 5. Change of averaged lead shock speed as a function of coal particle concentration for different particle sizes.

Plotted in Fig. 5 is the change of averaged lead shock speed, $\bar{D}$, as a function of particle concentrations for different particle diameters. It is calculated from the length of two-phase section (i.e., 0.1 m) divided by the total shock propagation time in this section. One can see that, for a fixed particle size, $\bar{D}$ gradually decreases when $c$ is increased. This indicates that coal particles have an inhibitory effect on methane detonation propagation, because the coal particles can absorb energy to heat themselves for the subsequent surface reactions. However, for low particle concentration, e.g., 10 and 50 g/m$^3$, $\bar{D}$ is slightly higher than that of the purely gaseous case ($c$ = 0, the pink square). This means that in a coal dust suspension with small particle concentrations, the speed can be slightly enhanced (by 2%) and leads to more stable DW (e.g., Fig. 3b). This may be because of the pressure impulse emanated from particle surface reactions behind the lead shock, as mentioned in Fig. 3 and will be discussed in Section 4.2. For 2.5 µm coal particles, detonation extinction without re-initiation occurs for high particle concentrations, i.e., $c = 500\text{-}1000$ g/m$^3$. Therefore, the average shock speed (open circles in Fig. 5) is generally lower than those of other cases. Nonetheless, for 1 µm particles with the same particle concentrations, due to DW re-initiation after extinction, the average shock speed is higher than those of 2.5 µm, as shown in Fig. 5.

*4.2 Hybrid detonation structure*

Detailed structures of methane/coal particle hybrid detonation will be analyzed in this section. The Eulerian gas and Lagrangian particle results with $c = 50$ g/m$^3$ and $d_p = 1$ µm (same as Fig. 3c) are demonstrated in Fig. 6. A weakly unstable detonation wave is observed. The Mach stem, incident wave, transverse wave and triple point can be identified, as annotated in Fig. 6(b). The gas reaction heat release rate $\dot{Q}$ is high immediately behind the lead shock front (LSF), as shown in Fig. 6(d). Moreover, unburned gas (UBG) pockets exist in the detonation products (see Figs. 6a and 6b), which are leaked behind the incident wave.

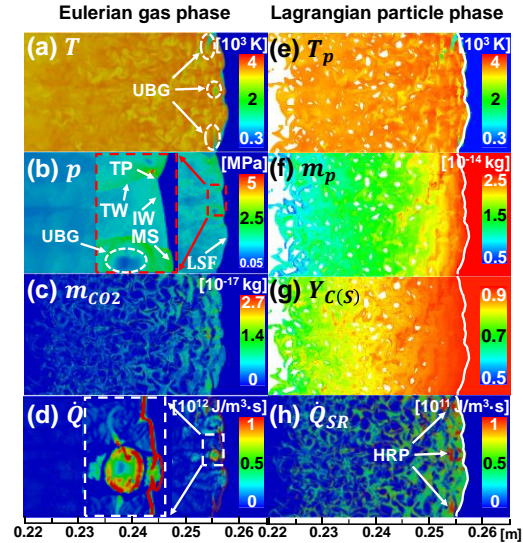

Fig. 6. Distributions of (a) gas temperature, (b) pressure, (c) $CO_2$ mass from surface reaction, (d) gas reaction heat release rate, (e) particle temperature, (f) particle mass, (g) carbon mass fraction in the particle, (h) char combustion heat release rate. $c = 50$ g/m$^3$ and $d_p = 1$ µm. MS: Mach stem, TP: triple point, IW: incident wave, TW: transverse wave, UBG: unburned gas, HRP: heat release point. White line: shock front.

One can see from Fig. 6(e) that the coal particles are heated to above 3,000 K immediately behind the LSF. This is reasonable because fine particles of 1 µm diameter have short thermal relaxation timescale. Accordingly, the char starts to burn, and high concentration $CO_2$ is produced from coal surface reaction behind Mach stems and incident waves (Fig. 6c). This leads to quick reduction of coal particle mass $m_p$, evidenced in Fig. 6(f). Within 0.01 m behind the LSF, the mass of most particles is reduced to around 50% of the original value. In Fig. 6(g), the carbon mass fraction in the particles, $Y_{C(S)}$, is reduced to approximately 70% (not burned out yet) at 0.02 m behind the LSF. Striped distributions of surface reaction from char combustion $\dot{Q}_{SR}$ can be found in Fig. 6(h). Several locations with high $\dot{Q}_{SR}$ can be seen (marked as HRP), which are caused by enhanced char combustion facilitated by the availability of the oxidant species in the unburned gas pockets. The localized strong surface reaction heat generation further promotes the homogeneous gas reactions, thereby higher $\dot{Q}$ near there (see Fig. 6d inset), which further elevates the local pressure. These pockets with char burning would be conducive for pressure wave formation, thereby affecting the lead shock.

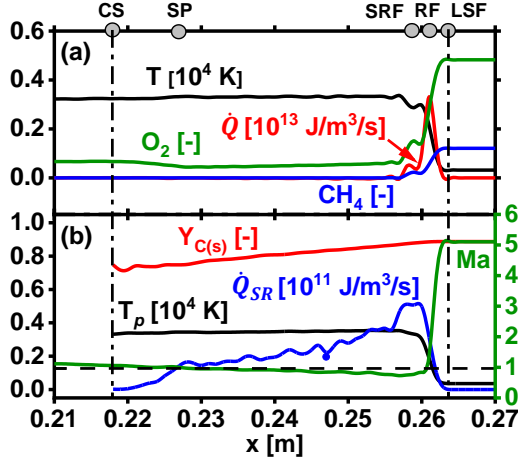

Fig. 7. Distributions of averaged (a) gas phase variable and (b) particle phase variable corresponding to the results in Fig. 6. $c = 50$ g/m$^3$ and $d_p = 1$ µm. LSF: lead shock front; RF: gas reaction front; SRF: surface reaction front; SP: shock-frame sonic point; CS: two-phase contact surface.

The mean structure of the hybrid detonation can be quantified through averaging the key gas (density-weighted averaging) and particle (simple averaging) variables along the domain width and the results are presented in Fig. 7. The length of the particle-laden area behind the lead shock is about 0.046 m, and the end of this area is a multiphase contact surface (CS). As observed from Fig. 7(a), the gas reaction HRR increases quickly after the shock and peaks around 0.26 m (termed as reaction front, RF). As such, the averaged induction distance between LSF and RF is about 3 mm. Accordingly, the mass fractions of $CH_4$ and $O_2$ quickly drop to around 0 and 0.045 respectively behind the reaction zone. The residual $O_2$ provides favorable environment for char combustion. The gas temperature $T$ rises rapidly to over 3,000 K due to detonative combustion, and the particle temperature $T_p$ basically follows the gas one due to the fast heating process. The maximum char combustion HRR (the corresponding location termed as SRF) lies slightly behind the RF. Nonetheless, continuous combustion of the coal particles leads to distributed char combustion HRR $\dot{Q}_{SR}$ in the detonated products. From the distributions of the shock-frame Mach number $Ma$, the subsonic (very close to the sonic condition, like a CJ detonation) region spans from $x = 0.23$ to 0.26 m. The location of $Ma = 1$ corresponds to the sonic point (SP). Therefore, char combustion largely proceeds in the subsonic region, which enables the influence of forward-running pressure waves from char combustion heat release on the lead shocks. The skeletal structure of the hybrid detonation is also marked along the top $x$-axis in Fig. 7.

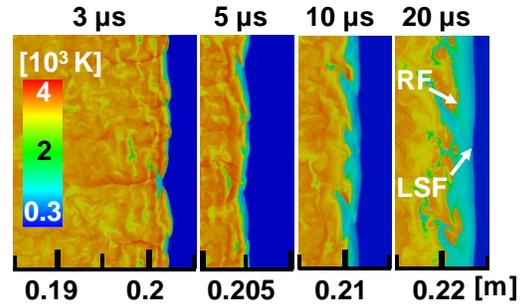

Fig.8. Time sequence of gas temperature in a detonation extinction. Tick spacing: 5 mm. $c = 1000$ g/m$^3$, $d_p = 1$ µm.

*4.3 Detonation extinction and re-initiation*

It has been shown from Fig. 5 that when the coal particle diameter is small (e.g., 1 µm) and concentration is high, detonation extinction and re-initiation occur. To elaborate on this unsteady process, the results correspond to Fig. 3(f), i.e., $c = 1000$ g/m$^3$ and $d_p = 1$ µm, will be discussed here. Figure 8 shows the time evolutions of gas temperature at four instants. At 3 µs, when the DW just enters the two-phase section starting at $x = 0.2$ m, it is weakly unstable with multiple heads. However, at 5 µs, the distance between the LSF and RF generally increases, and the post-shock temperature is reduced to below 3,100 K. Afterwards, the distance is further increased, and the detonative combustion extinguishes. This indicates that considerably energy is extracted from the gas to heat the coal particles and hence coupling between the shock and reaction front cannot maintain.

After the shock wave propagates a distance, re-detonation occurs. Three instants of this transient are shown in Fig. 9. Note that they are the continued development of the results in Fig. 8. At 30 μs, most of the LSF and RF are still decoupled. However, some evolving hot spots of different sizes appear along the RF, which are numbered as 1-3 in Fig. 9. They are characterized by locally elevated pressure (see the red colour in Fig. 9b), indicating the nature of isochoric combustion caused by the coherent interplay between strong heat release and pressure waves.

The flow structure behind the lead shock can be clearly found from the pressure gradient magnitude in Fig. 9(c). The arched and interactive shocks are observable, which originate from spatially nonuniform char combustion heat release from dispersed particles in the post-shock subsonic zones (see the shock-frame sonic isolines, $Ma = 1$). We also did a test with surface reaction de-activated for this case: there are no curved shocks behind the lead shock, and no re-detonation occurs (see supplementary document). The propagation of these shocks results in: (1) the forward-running components overtake and hence intensify the LSF; (2) the spanwise components re-compress the shocked gas and coal particles behind the LSF; and (3) more importantly, shock-focusing along the RF by these shocks leads to the formation of small reactive spots (e.g., 2 and 3). These spots quickly grow logitudinally and spanwise in the form of propagating reaction fronts, as evidenced in the results of 31 and 32 μs. Their leading sections intersect with the LSF, which generates a overdriven Mach stem with strong gas reaction HRR. The spanwise component evlves into the transverse wave extending from the triple pionts of the new MS (see 32 μs results). As such, the initial number of DW heads is correlated to the number of the hot spot and therefore randomness exists. This randomness largely comes from the inducing factors for hot spot formation, e.g., heterogeneous reaction, shock focusing location, and chemistry-shock interaction.

A regime map for detonation extinction and re-initiation with different coal particle concentrations is predicted in Fig. 10. Here $d_p = 1$ μm. The critical extinction (re-detonation) location is determined from the *x*-coordinate where the peak pressure is critically lower than (exceeds) 2.5 MPa, as shown with the two dashed lines in Fig. 3(e). It is seen that methane detonation extinction and re-detonation only occur when $c > 465$ g/m$^3$. The critical extinction location is not sensitive to the coal particle concentration. It is close to 0.2 m, indicating that extinction occurs almost immediately when the DW arrives at the particle suspensions. Nonetheless, the critical re-initiation location decreases with particle concentration. This is reasonable because higher concentration of coal particles leads to greater interphase exchanges of momentum and (mainly) energy. However, as the particle concentration exceeds 1000 g/m$^3$, the re-initiation location approaches a constant value of 0.225 m. This may be limited by the timescales of coal particle heating and/or char combustion.

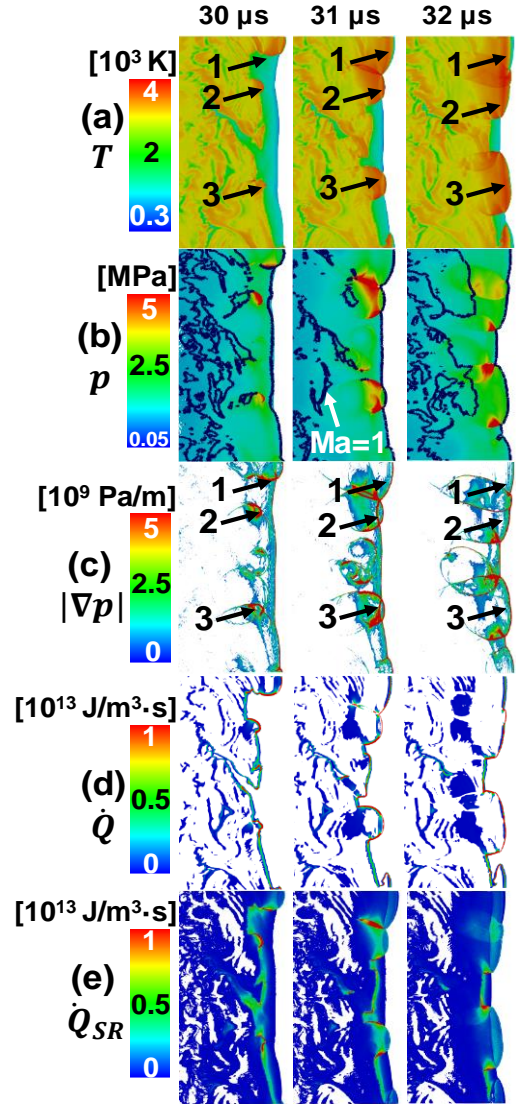

Fig. 9. Distributions of (a) gas temperature, (b) pressure, (c) pressure gradient magnitude, (d) gas reaction HRR, and (e) surface reaction HRR in a re-detonation process. 1, 2 and 3: hot spots. $c = 1000$ g/m$^3$ and $d_p = 1$ μm.

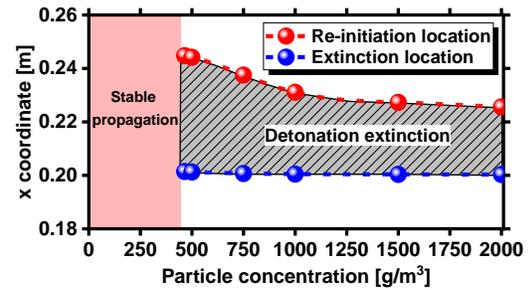

Fig. 10. Regime map of extinction and re-detonation with different coal particle concentrations. $d_p = 1$ μm.

## 5 Conclusions

Methane detonation extinction and re-initiation in dilute coal particle suspensions are simularted with Eulerian-Lagrangian appraoch. The results show that the methane detonation wave propagation is considerably affected by coal particle concentration and size. Detonation extinction occurs when the particle size is small and concentration is high. The averaged lead shock speed decreases with increased particle concentration and decreased particle size. Mean structure of the methane/coal particle hybrid detonation is analysed, based on the gas and particle quantities. Moreover, for 1 μm particle, if the particle concentration is beyond a threshold value, detonation re-initiation occurs. This is caused by the shock focusing along the reaction front in a decoupled detonation and these shocks are generated from char combustion behind the lead shock. A regime map of detonation propagation and extinction is predicted. It is found that the re-initiation location decreases with the particle concentration and approaches a constant value when the concentration exceeds 1000 g/m$^3$.

## Acknowledgements

This work used the Fugaku Supercomputer in Japan (hp210196). HZ is supported by MOE Tier 1 Grant (R-265-000-653-114). WR is supported by the Fundamental Research Funds for the Central Universities (2019XKQYMS75). JS is funded by the CSC Scholarship (202006420042).

## References

[1] A.A. Vasil'ev, A. V Pinaev, A.A. Trubitsyn, A.Y. Grachev, A. V Trotsyuk, P.A. Fomin, A. V Trilis, What is burning in coal mines: Methane or coal dust?, Combust. Explos. *Shock Waves*. 53 (2017) 8–14.

[2] C.T. Cloney, R.C. Ripley, M.J. Pegg, P.R. Amyotte, Laminar combustion regimes for hybrid mixtures of coal dust with methane gas below the gas lower flammability limit, *Combust. Flame*. 198 (2018) 14–23.

[3] H. Xu, X. Wang, R. Gu, H. Zhang, Experimental study on characteristics of methane–coal-dust mixture explosion and its mitigation by ultra-fine water mist, *J. Eng. Gas Turbines Power*. 134 (2012) 061401.

[4] H. Xu, X. Wang, Y. Li, P. Zhu, H. Cong, W. Qin, Experimental investigation of methane/coal dust explosion under influence of obstacles and ultrafine water mist, *J. Loss Prev. Process Ind*. 49 (2017) 929–937.

[5] Y. Xie, V. Raghavan, A.S. Rangwala, Study of interaction of entrained coal dust particles in lean methane–air premixed flames, *Combust. Flame*. 159 (2012) 2449–2456.

[6] S.R. Rockwell, A.S. Rangwala, Influence of coal dust on premixed turbulent methane–air flames, *Combust. Flame*. 160 (2013) 635–640.

[7] D. Chen, *Study on Propagation Characteristics and Mechanism of Methane/Coal Dust Mixture Flame,* PhD thesis. University of Science and Technology of China, Hefei, China, 2007.

[8] P.R. Amyotte, K.J. Mintz, M.J. Pegg, Y.H. Sun, The ignitability of coal dust-air and methane-coal dust-air mixtures, *Fuel*. 72 (1993) 671–679.

[9] R.W. Houim, E.S. Oran, Numerical simulation of dilute and dense layered coal-dust explosions, *Proc. Combust. Inst*.35 (2015) 2083-2090.

[10] R.W. Houim, E.S. Oran, Structure and flame speed of dilute and dense layered coal-dust explosions, *J. Loss Prev. Process Ind.* (2015) 214–222.

[11] S. Guhathakurta, R.W. Houim, Influence of thermal radiation on layered dust explosions, *J. Loss Prev. Process Ind.*. 72 (2021) 104509.

[12] K. Shimura, A. Matsuo, Using an extended CFD–DEM for the two-dimensional simulation of shock-induced layered coal-dust combustion in a narrow channel, *Proc. Combust. Inst*. 37 (2019) 3677–3684.

[13] C.T. Crowe, J.D. Schwarzkopf, M. Sommerfeld, Y. Tsuji, *Multiphase flows with droplets and particles,* CRC Press, New York, U.S., 1998, p. 29.

[14] J. Shi, Y. Xu, W. Ren, H. Zhang, Critical condition and transient evolution of methane detonation extinction by fine water droplet curtains, *Fuel*. (2022), in press.

[15] A. Kazakov and M. Frenklach, Reduced Reaction Sets based on GRI-Mech 1.2,. <http://combustion.berkeley.edu/drm/>.

[16] L. Qiao, Transient flame propagation process and flame-speed oscillation phenomenon in a carbon dust cloud, *Combust. Flame*. 159 (2012) 673–685.

[17] M.M. Baum, P.J. Street, Predicting the combustion behaviour of coal particles, *Combust. Sci. Technol*. 3 (1971) 231–243.

[18] M. Stöllinger, B. Naud, D. Roekaerts, N. Beishuizen, S. Heinz, PDF modeling and simulations of pulverized coal combustion – Part 2: Application, *Combust. Flame*. 160 (2013) 396–410.

[19] X. Zhao, D.C. Haworth, Transported PDF modeling of pulverized coal jet flames, *Combust. Flame*. 161 (2014) 1866–1882.

[20] Z. Huang, H. Zhang, On the interactions between a propagating shock wave and evaporating water droplets, *Phys. Fluids*. 32 (2020) 123315.

[21] Z. Huang, M. Zhao, Y. Xu, G. Li, H. Zhang, Eulerian-Lagrangian modelling of detonative combustion in two-phase gas-droplet mixtures with OpenFOAM: Validations and verifications, *Fuel*. 286 (2021) 119402.

[22] H. Zhang, M. Zhao, Z. Huang, Large eddy simulation of turbulent supersonic hydrogen flames with OpenFOAM, *Fuel*. 282 (2020) 118812.

[23] M. Zhao, Z. Ren, H. Zhang, Pulsating detonative combustion in n-heptane/air mixtures under off-stoichiometric conditions, *Combust. Flame*. 226 (2021) 285–301.

[24] M. Zhao, M.J. Cleary, H. Zhang, Combustion mode and wave multiplicity in rotating detonative combustion with separate reactant injection, *Combust. Flame*. 225 (2021) 291–304.

[25] A. Kurganov, S. Noelle, G. Petrova, Semidiscrete central-upwind schemes for hyperbolic conservation laws and Hamilton--Jacobi equations, *SIAM J. Sci. Comput*. 23 (2001) 707–740.

[26] Y. Xu, M. Zhao, H. Zhang, Extinction of incident hydrogen/air detonation in fine water sprays, *Phys. Fluids*. 33 (2021) 116109.

[27] M. Sommerfeld, The unsteadiness of shock waves propagating through gas-particle mixtures, *Exp. Fluids* 3 (1985) 197–206.

[28] B. Veyssiere, Detonations in gas-particle mixtures, *J. Propuls. Power*. 22 (2006) 1269–1288.